\documentclass[10pt,letterpaper]{article}
\usepackage{opex3}
\usepackage{cite}

\begin{document}

\title{Generation of three-dimensional entangled state between a single atom
and a Bose-Einstein condensate via adiabatic passage}

\author{Li-Bo Chen,$^{1,2}$ Peng Shi,$^{1}$ Chun-Hong Zheng,$^{1,2}$ and
Yong-Jian Gu $^{1,*}$}

\address{$^{1}$Department of Physics, Ocean University of China, Qingdao 266100, China \\ $^{2}$School of Science, Qingdao Technological
University, Qingdao 266033, China}

\email{$^{*}$yjgu@ouc.edu.cn} 



\begin{abstract}
Inspired by a recently experiment by M. Lettner \emph{et al}. [Phys. Rev. Lett. \textbf{106}, 210503 (2011)], we propose a robust scheme to prepare three-dimensional entanglement state
between a single atom and a Bose-Einstein condensate (BEC) via stimulated
Raman adiabatic passage (STIRAP) technique. The atomic spontaneous
radiation, the cavity decay, and the fiber loss are efficiently suppressed
by the engineering adiabatic passage. Our strictly numerical simulation shows
our proposal is good enough to demonstrate the generation of three-dimensional entanglement with high fidelity and within the current experimental technology.
\end{abstract}



\section{Introduction}

Quantum entanglement plays a vital role in many practical quantum
information system, such as quantum teleportation \cite{bennett}, quantum
dense coding \cite{bennett2}, and quantum cryptography \cite{Ekert}.
Entangled states of higher-dimensional systems are of great interest owing
to the extended possibilities they provide, which including higher
information density coding \cite{Durt}, stronger violations of local realism
\cite{Kaszlikowski,Collins}, and more resilience to error \cite{Fujiwara}
than two dimensional system. Over the past few years, fairish attention has
been paid to implement higher-dimensional entanglement with trapped ions
\cite{Klimov,Linington}, photons \cite{Vaziri,Lanyon,Dada}, and cavity QED \cite{Zou,Lin,saiyun,libo1}.

Moreover, it has been shown that entanglement between two spatially separated subsystems is very useful for distributed quantum computation \cite{bennett3,kimble}. Recently, a large number of schemes have been proposed for generating entangled state of atoms, which are individually trapped in distant optical cavities connected by fibers \cite{cirac,tpe,sjvan,sclark,a.s,yin,lu}. The main problems in entangling atoms in these schemes are the decoherence due to leakage of photons from the cavity and fiber modes, and spontaneous radiation of the atoms \cite{Mabuchi}. By using the stimulated Raman adiabatic passage (STIRAP) \cite{Oreg,Gaubatz,Gaubatz2,Bergmann,kuk,Unanyan1,Theuer,song}, our scheme can overcome these
problems. The idea of STIRAP is that the system is initially prepared in a
decoherence-free state (dark state), and evolve adiabatically along the dark
state to the required state by two delayed but partially overlapping pulses.
Many schemes have been proposed to prepare entanglement state via STIRAP \cite{Unanyan2,Unanyan3,ma,ma1,nv,kis,songjie,Klein,libo}.

The Bose-Einstein condensate (BEC) has many advantages over other systems such as long storage times, the high write-read efficiencies, and excellent internal-state preparation \cite{Yoshikawa,Riedl}.
Recently, remote entanglement between a single atom and BEC was experimentally realized \cite{Lettner}. But the
efficiency is very low due to the photon loss. In this paper, we takes both
the advantages of cavity-fiber system and STIRAP in order to create
three-dimensional entanglement state between a single $^{87}Rb$ atom and a $^{87}Rb$\ BEC at a distance. The atom and BEC are placed inside two high-finesse
optical cavities respectively, which connected by an optical fibre. The atom--light interaction is identical for all atoms of the BEC and enhanced greatly because the atoms collectively couple to the same light mode \cite{Brennecke,Klaers}. The entanglement state can be generated with highly fidelity even in the range that the cavity decay and spontaneous radiation of the atoms are comparable with the atom-cavity coupling strength. Our scheme is also robust to the variation of atom number in the BEC. As a result, the highly fidelity three-dimensional entanglement state of the BEC and atom can be realized base on our proposed scheme.

This paper is organized as follows. In Sec. 2, we introduce the basic model of our system. In Sec. 3, the generation of the three-dimensional entanglement state is provided. In Sec. 4, we demonstrate the influences of atomic spontaneous radiation, photon leakage out of the cavities and fiber on the implementation. Finally, in Sec. 5, we discuss experimental feasibility of our scheme and conclude our results.
\section{The fundamental model}
We consider the situation describe in Fig. 1, where a single $^{87}Rb$ atom
and a $^{87}Rb$ BEC are trapped in two distant double-mode optical cavities,
which are connected by an optical fiber (see Fig. 1). The $^{87}Rb$ atomic
levels and transitions are also depicted in this Fig. \cite{Lettner,wilk,Weber}.
The states $\left\vert g_{L}\right\rangle $, $\left\vert g_{0}\right\rangle $%
, $\left\vert g_{R}\right\rangle $ and $\left\vert g_{a}\right\rangle $
correspond to $\left\vert F=1,m_{F}=-1\right\rangle $, $\left\vert
F=1,m_{F}=0\right\rangle $, $\left\vert F=1,m_{F}=1\right\rangle $ of $%
5S_{1/2}$ and $\left\vert F=2,m_{F}=0\right\rangle $ of $5S_{1/2}$, while $%
\left\vert e_{L}\right\rangle $, $\left\vert e_{0}\right\rangle $ and $%
\left\vert e_{R}\right\rangle $ correspond to $\left\vert
F=1,m_{F}=-1\right\rangle $, $\left\vert F=1,m_{F}=0\right\rangle $ and $%
\left\vert F=1,m_{F}=1\right\rangle $ of $5P_{3/2}$.
The atomic transition $\left\vert g_{a}\right\rangle \leftrightarrow
\left\vert e_{0}\right\rangle $ of atom in cavity $A$ is driven resonantly
by a $\pi $-polarized classical field with Rabi frequency $\Omega _{A}$; $%
\left\vert e_{0}\right\rangle _{A}\leftrightarrow \left\vert
g_{L}\right\rangle _{A}$ $\left( \left\vert e_{0}\right\rangle
_{A}\leftrightarrow \left\vert g_{R}\right\rangle _{A}\right) $ is
resonantly coupled to the cavity mode $a_{A,L}$ $\left( a_{A,R}\right) $ with
coupling constant $g_{A}$.
The atomic transition $\left\vert g_{L}\right\rangle _{B}\leftrightarrow
\left\vert e_{L}\right\rangle _{B}$ $\left( \left\vert g_{R}\right\rangle
_{B}\leftrightarrow \left\vert e_{R}\right\rangle _{B}\right) $ of BEC in
cavity $B$\ is driven resonantly by a $\pi $-polarized classical field with
Rabi frequency $\Omega _{B}$; $\left\vert e_{R}\right\rangle
_{B}\leftrightarrow \left\vert g_{0}\right\rangle _{B}$ $\left( \left\vert
e_{L}\right\rangle _{B}\leftrightarrow \left\vert g_{0}\right\rangle
_{B}\right) $ is resonantly coupled to the cavity mode $a_{B,L}$ $\left(
a_{B,R}\right) $ with coupling constant $g_{B}$.
Here we consider BEC for a single excitation, the ground and single excitation states are described by the state vectors $\left\vert G_{f}\right\rangle =\left( 1/%
\sqrt{N}\right) \sum_{j=1}^{N}\left\vert g_{f}\right\rangle _{j}\otimes
_{k=1,k\neq j}^{N}\left\vert g_{0}\right\rangle _{j}$ and $\left\vert
E_{f}\right\rangle =\left( 1/\sqrt{N}\right) \sum_{j=1}^{N}\left\vert
e_{f}\right\rangle _{j}\otimes _{k=1,k\neq j}^{N}\left\vert
g_{0}\right\rangle _{j}$ $\left( f=0, L, R\right) $, where $\left\vert
...\right\rangle _{j}$ describe the state of the $j$th atom in the BEC \cite
{Lettner}.
\begin{figure}[htbp]
\centering\includegraphics[width=9cm]{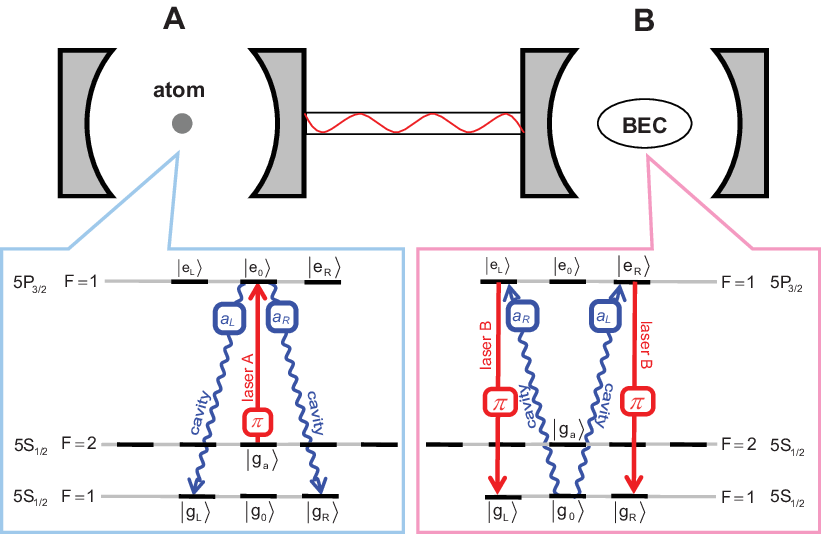}
\caption{A single $^{87}Rb$ atom and a $^{87}Rb$ BEC are trapped in two
distant double-mode optical cavities, which are connected by an optical
fiber. The states $\left\vert g_{L}\right\rangle $, $\left\vert
g_{0}\right\rangle $, $\left\vert g_{R}\right\rangle $ and $\left\vert
g_{a}\right\rangle $ correspond to $\left\vert F=1\ ,%
m_{F}=-1\right\rangle $, $\left\vert F=1\ ,m_{F}=0\right\rangle $, $%
\left\vert F=1\ ,m_{F}=1\right\rangle $ of $5S_{1/2}$ and $\left\vert
F=2\ ,m_{F}=0\right\rangle $ of $5S_{1/2}$, while $\left\vert
e_{L}\right\rangle $, $\left\vert e_{0}\right\rangle $ and $\left\vert
e_{R}\right\rangle $ correspond to $\left\vert F=1\ ,%
m_{F}=-1\right\rangle $, $\left\vert F=1\ ,m_{F}=0\right\rangle $ and $%
\left\vert F=1\ ,m_{F}=1\right\rangle $ of $5P_{3/2}$. The atomic
transition $\left\vert g_{a}\right\rangle \leftrightarrow\left\vert
e_{0}\right\rangle $ of atom in cavity $A$ is driven resonantly by a $\pi$%
-polarized classical field with Rabi frequency $\Omega_{A}$; $\left\vert
e_{0}\right\rangle _{A}\leftrightarrow \left\vert g_{L}\right\rangle _{A}$ $%
\left( \left\vert e_{0}\right\rangle _{A}\leftrightarrow\left\vert
g_{R}\right\rangle _{A}\right) $ is resonantly coupled to the cavity mode $%
a_{A,L}$ $\left( a_{A,R}\right) $ with coupling constant $g_{A}$. The atomic
transition $\left\vert g_{L}\right\rangle _{B}\leftrightarrow\left\vert
e_{L}\right\rangle _{B}$ $\left( \left\vert g_{R}\right\rangle
_{B}\leftrightarrow\left\vert e_{R}\right\rangle _{B}\right) $ of BEC in
cavity $B$\ is driven resonantly by a $\pi$-polarized classical field with
Rabi frequency $\Omega_{B}$; $\left\vert e_{R}\right\rangle
_{B}\leftrightarrow\left\vert g_{0}\right\rangle _{B}$ $\left( \left\vert
e_{L}\right\rangle _{B}\leftrightarrow\left\vert g_{0}\right\rangle
_{B}\right) $ is resonantly coupled to the cavity mode $a_{B,L}$ $\left(
a_{B,R}\right) $ with coupling constant $g_{B}$.}
\end{figure}

Initially, if the atom and BEC are prepared in the states $\left\vert g_{a}\right\rangle _{A}$ and $\left\vert G_{0}\right\rangle _{B}$
respectively, and the cavities and fiber modes are in the vacuum states. In the rotating
wave approximation, the interaction Hamiltonian of the atom (BEC)-cavity system can
be written as (setting $\hbar =1$) \cite{Brennecke}
\begin{eqnarray}
H_{ac}&&=\sum_{k=L,R}(\Omega _{A}(t)\left\vert e_{0}\right\rangle
_{A}\left\langle g_{a}\right\vert +g_{A}(t)a_{A,k}\left\vert
e_{0}\right\rangle _{A}\left\langle g_{k}\right\vert
+\sqrt{N}\Omega _{B}(t)\left\vert E_{k}\right\rangle _{B}\left\langle
G_{k}\right\vert) \nonumber \\ &&+\sqrt{N}g_{B}(t)a_{B,L}\left\vert E_{R}\right\rangle
_{B}\left\langle G_{0}\right\vert +\sqrt{N}g_{B}(t)a_{B,R}\left\vert E_{L}\right\rangle
_{B}\left\langle G_{0}\right\vert+H.c.,  \label{1}
\end{eqnarray}%
In the short fibre limit, the coupling between the cavity fields and the
fiber modes can be written as the interaction Hamiltonian \cite{tpe,a.s,yin}
\begin{equation}
H_{cf}=\sum_{k=L,R}\nu _{k}\left[ b_{k}\left( a_{A,k}^{+}+a_{B,k}^{+}\right)
+H.c.\right] .  \label{2}
\end{equation}%
In the interaction picture the total Hamiltonian now becomes%
\begin{equation}
H_{I}=H_{ac}+H_{cf}.  \label{3}
\end{equation}%

\section{Generation of the three-dimensional entanglement state}

In this section, we begin to investigate the generation of the
three-dimensional entangled state in detail. The time evolution of the whole
system state is governed by the Schr\"{o}dinger equation
\begin{equation}
i\frac{\partial}{\partial t}\left\vert \psi\left( t\right) \right\rangle
=H_{I}\left\vert \psi\left( t\right) \right\rangle .\  \label{4}
\end{equation}
 The single excitation subspace can be spanned by the following state
vectors \cite{zheng}
\begin{eqnarray}
\left\vert \phi_{1}\right\rangle & =\left\vert g_{a}\right\rangle
_{A}\left\vert G_{0}\right\rangle _{B}\left\vert 0000\right\rangle
_{c}\left\vert 00\right\rangle _{f},\  \nonumber \\
\left\vert \phi_{2}\right\rangle & =\left\vert e_{0}\right\rangle
_{A}\left\vert G_{0}\right\rangle _{B}\left\vert 0000\right\rangle
_{c}\left\vert 00\right\rangle _{f},\  \nonumber \\
\left\vert \phi_{3}\right\rangle & =\left\vert g_{L}\right\rangle
_{A}\left\vert G_{0}\right\rangle _{B}\left\vert 1000\right\rangle
_{c}\left\vert 00\right\rangle _{f},\  \nonumber \\
\left\vert \phi_{4}\right\rangle & =\left\vert g_{R}\right\rangle
_{A}\left\vert G_{0}\right\rangle _{B}\left\vert 0100\right\rangle
_{c}\left\vert 00\right\rangle _{f},\  \nonumber \\
\left\vert \phi_{5}\right\rangle & =\left\vert g_{L}\right\rangle
_{A}\left\vert G_{0}\right\rangle _{B}\left\vert 0000\right\rangle
_{c}\left\vert 10\right\rangle _{f},\  \nonumber \\
\left\vert \phi_{6}\right\rangle & =\left\vert g_{R}\right\rangle
_{A}\left\vert G_{0}\right\rangle _{B}\left\vert 0000\right\rangle
_{c}\left\vert 01\right\rangle _{f},\  \nonumber \\
\left\vert \phi_{7}\right\rangle & =\left\vert g_{L}\right\rangle
_{A}\left\vert G_{0}\right\rangle _{B}\left\vert 0010\right\rangle
_{c}\left\vert 00\right\rangle _{f},\  \nonumber \\
\left\vert \phi_{8}\right\rangle & =\left\vert g_{R}\right\rangle
_{A}\left\vert G_{0}\right\rangle _{B}\left\vert 0001\right\rangle
_{c}\left\vert 00\right\rangle _{f},\ \nonumber \\
\left\vert \phi_{9}\right\rangle & =\left\vert g_{L}\right\rangle
_{A}\left\vert E_{R}\right\rangle _{B}\left\vert 0000\right\rangle
_{c}\left\vert 00\right\rangle _{f},\  \nonumber \\
\left\vert \phi_{10}\right\rangle & =\left\vert g_{R}\right\rangle
_{A}\left\vert E_{L}\right\rangle _{B}\left\vert 0000\right\rangle
_{c}\left\vert 00\right\rangle _{f},\ \nonumber \\
\left\vert \phi_{11}\right\rangle & =\left\vert g_{L}\right\rangle
_{A}\left\vert G_{R}\right\rangle _{B}\left\vert 0000\right\rangle
_{c}\left\vert 00\right\rangle _{f},\  \nonumber \\
\left\vert \phi_{12}\right\rangle & =\left\vert g_{R}\right\rangle
_{A}\left\vert G_{L}\right\rangle _{B}\left\vert 0000\right\rangle
_{c}\left\vert 00\right\rangle _{f},\  \label{5}
\end{eqnarray}
where $\left\vert n_{AL}\,n_{AR}\,n_{BL}\,%
n_{BR}\right\rangle _{c}$ denotes the field state with $n_{Ai}$ ($i=L$, $R$)
photons in the $i$ polarized mode of cavity $A$, $n_{Bi}$ in the $i$
polarized mode of cavity $B$, and $\left\vert n_{L}\,%
n_{R}\right\rangle _{f}$ represents $n_{i}$ photons in $i$ polarized mode of
the fiber. The Hamiltonian $H_{I}$ has the following dark state:
\begin{eqnarray}
\left\vert D(t)\right\rangle  =&&K\{2g_{A}(t)\Omega_{B}(t)\left\vert \phi
_{1}\right\rangle -\Omega_{A}(t)\Omega_{B}(t)\left[ \left\vert \phi
_{3}\right\rangle +\left\vert \phi_{4}\right\rangle -\left\vert \phi
_{7}\right\rangle -\left\vert \phi_{8}\right\rangle \right]  \nonumber \\
&& -g_{B}(t)\Omega_{A}(t)\left[ \left\vert \phi_{11}\right\rangle +\left\vert
\phi_{12}\right\rangle \right] \},\  \label{6}
\end{eqnarray}
which is the eigenstate of the Hamiltonian corresponding to zero eigenvalue.
Here and in the following $g_{i}$, $\Omega_{i}$ are real, and $%
K^{-2}=g_{A}^{2}\Omega_{B}^{2}+4\Omega_{A}^{2}\Omega_{B}^{2}+2g_{B}^{2}%
\Omega _{A}^{2}$. Under the condition
\begin{equation}
g_{A}(t) ,\ g_{B}(t)\gg\Omega_{A}(t) ,\ \Omega_{B}(t),\
\label{7}
\end{equation}
we have%
\begin{equation}
\left\vert D(t)\right\rangle \sim 2g_{A}(t)\Omega_{B}(t)\left\vert \phi
_{1}\right\rangle -g_{B}(t)\Omega_{A}(t)\left[ \left\vert
\phi_{11}\right\rangle +\left\vert \phi_{12}\right\rangle \right] .\
\label{8}
\end{equation}
Suppose the initial state of the system is $\left\vert \phi_{1}\right\rangle
$, if we design pulse shapes such that
\begin{eqnarray}
\lim_{t\rightarrow-\infty}\frac{g_{B}(t)\Omega_{A}(t)}{g_{A}(t)\Omega_{B}(t)}
& =0,\  \nonumber \\
\lim_{t\rightarrow+\infty}\frac{g_{A}(t)\Omega_{B}(t)}{g_{B}(t)\Omega_{A}(t)}
& =\frac{1}{2},\  \label{9}
\end{eqnarray}
we can adiabatically transfer the initial state $\left\vert \phi
_{1}\right\rangle $ to a equal superposition of $\left\vert \phi_{1}\right\rangle $%
, $\left\vert \phi_{11}\right\rangle $ and $\left\vert
\phi_{12}\right\rangle $, i.e., $1/\sqrt{3}(\left\vert g_{a}\right\rangle
_{A}\left\vert G_{0}\right\rangle _{B}-\left\vert g_{L}\right\rangle
_{A}\left\vert G_{R}\right\rangle _{B}-\left\vert g_{R}\right\rangle
_{A}\left\vert G_{L}\right\rangle _{B})\left\vert 0000\right\rangle
_{c}\left\vert 00\right\rangle _{f}$, which is a product state of the
three-dimensional atom-BEC entangled state, the cavity mode vacuum state, and the
fiber mode vacuum state.
\bigskip The pulse shapes and sequence can be designed by an appropriate
choice of the parameters. The coupling rates are chosen such that $%
g_{A}(t)=g_{B}(t)=g$, $\nu_{L}=\nu_{R}=\nu=100g$, $N=10^{4}$, laser Rabi
frequencies are chosen as $\Omega_{A}(t)=\Omega_{0}\exp\left[ -\left(
t-t_{0}\right) ^{2}/200\tau^{2}\right] $ and $\Omega_{B}(t)=\Omega_{0}\exp%
\left[ -t^{2}/200\tau^{2}\right] +\frac{\Omega_{0}}{2}\exp\left[ -\left(
t-t_{0}\right) ^{2}/200\tau^{2}\right] $, with $t_{0}=20\tau$ being the
delay between pulses \cite{Kral}.
Fig. 2 shows the simulation results of the entanglement generation
process, where we choose $g=5\Omega_{0}$, $\tau=$ $\Omega_{0}^{-1}$. With this choice, conditions (7) and (8) can be well satisfied. The
Rabi frequencies of $\Omega_{A}(t)$, $\Omega_{B}(t)$ are shown in Fig. 2(a).
Fig. 2(b) and 2(c) shows the time evolution of populations. In Fig. 2(b) $%
P_{1}$, $P_{11}$, and $P_{12}$ denote the populations of the states $%
\left\vert \phi_{1}\right\rangle $, $\left\vert \phi_{11}\right\rangle $,
and $\left\vert \phi_{12}\right\rangle $. Fig. 2 (c) show the time evolution
of populations of other states $\{\left\vert \phi_{2}\right\rangle $, $%
\left\vert \phi _{3}\right\rangle $, $\left\vert \phi_{4}\right\rangle $, $%
\left\vert \phi _{5}\right\rangle $, $\left\vert \phi_{6}\right\rangle $, $%
\left\vert \phi _{7}\right\rangle $, $\left\vert \phi_{8}\right\rangle $, $%
\left\vert \phi _{9}\right\rangle $, $\left\vert \phi_{10}\right\rangle \}$,
which are almost zero during the whole dynamics. Finally $P_{1}$, $P_{11}$,
and $P_{12}$ arrive at $1/3$, which means the successful generation of the
3-dimensional entangled state. Figure 2(d) shows the error probability
defined by \cite{go}:
\begin{equation}
P_{e}\left( t\right) =1-\left\vert \left\langle D\left( t\right) \right\vert
\varphi_{s}\left( t\right) \rangle\right\vert ^{2},  \label{10}
\end{equation}
here $\left\vert \varphi_{s}\left( t\right) \right\rangle $ is the state
obtained by numerical simulation of Hamiltonian (3) and $\left\vert
D(t)\right\rangle $ is the dark state defined by Eq. (6). From the Fig.
2(a)-(d) we conclude that we can prepare the three-dimensional entanglement state between
single atom and a BEC with high success probability.
\begin{figure}[htbp]
\centering\includegraphics[width=7cm]{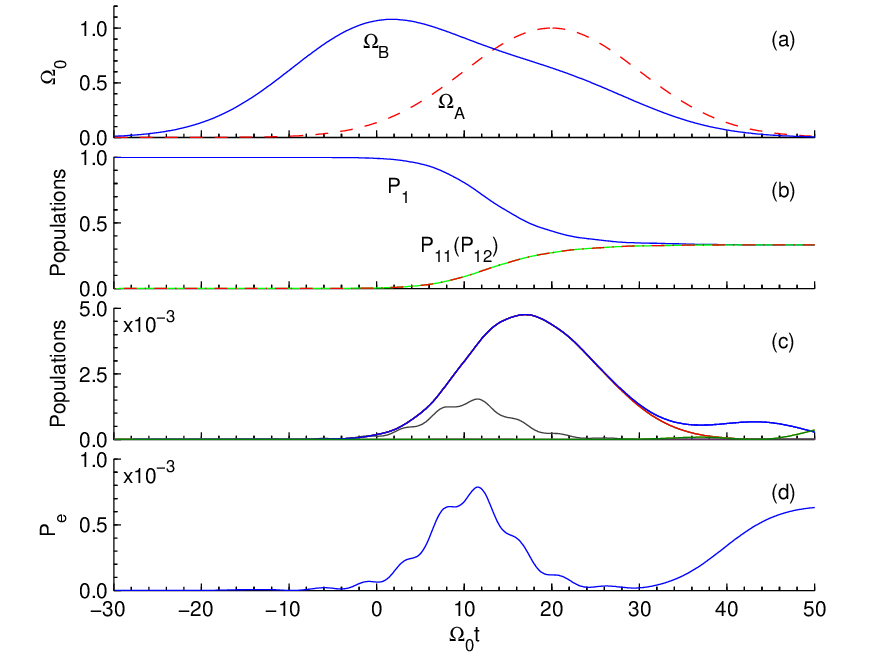}
\caption{The numerical simulation of Hamiltonian (3) in the entanglement
generation process, where we choose $g=5\Omega_{0}$, $\tau=$ $%
\Omega_{0}^{-1} $. (a): the Rabi frequency of $\Omega_{A}(t)$, $%
\Omega_{B}(t)$. (b): the time evolution of populations of the states $%
\left\vert \phi _{1}\right\rangle $, $\left\vert \phi_{11}\right\rangle $,
and $\left\vert \phi_{12}\right\rangle $ is denoted by $P_{1}$, $P_{11}$,
and $P_{12}$ respectively. (c): time evolution of populations of other
states $\{\left\vert \phi_{2}\right\rangle $, $\left\vert
\phi_{3}\right\rangle $, $\left\vert \phi_{4}\right\rangle $, $\left\vert
\phi_{5}\right\rangle $, $\left\vert \phi_{6}\right\rangle $, $\left\vert
\phi_{7}\right\rangle $, $\left\vert \phi_{8}\right\rangle $, $\left\vert
\phi_{9}\right\rangle $, $\left\vert \phi_{10}\right\rangle \}$ , which are almost
zero during the whole dynamics. (d): error probability $P_{e}\left(
t\right) $ defined by Eq. (6).}
\end{figure}

\section{Effects of spontaneous emission and photon leakage}

To evaluate the performance of our scheme, we now consider the dissipative
processes due to spontaneous decay of the atoms from the excited states and
the decay of cavity. We assess the effects through the numerical integration
of the master equation for the system in the Lindblad form. The master
equation for the density matrix of whole system can be expressed as \cite{lu}
\begin{eqnarray}
\frac{d\rho }{dt} =&&-i\left[ H_{I},\  \rho \right] -\sum_{k=L,R}[\frac{%
\kappa _{fk}}{2}\left( b_{k}^{+}b_{k}\rho -2b_{k}\rho
b_{k}^{+}+\rho b_{k}^{+}b_{k} \right)  \nonumber \\
&& -\sum_{i=A,B}\frac{\kappa _{ik}}{2}\left( a_{ik}^{+}a_{ik}\rho
-2a_{ik}^{+}\rho a_{ik}+\rho a_{ik}^{+}a_{ik}\right) ]  \nonumber \\
&& -\sum_{j=a,L,R}\frac{\gamma _{0j}^{A}}{2}\left( \sigma
_{e_{0}e_{0}}^{A}\rho -2\sigma _{g_{j}e_{0}}^{A}\rho \sigma
_{e_{0}g_{j}}^{A}+\rho \sigma _{e_{0}e_{0}}^{A}\right)  \nonumber \\
&& -\sum_{h=1}^{N}\sum_{k=L,R}\sum_{j=k,0}\frac{\gamma _{kj}^{Bh}}{2}\left( \sigma
_{e_{k}e_{k}}^{Bh}\rho -2\sigma _{g_{j}e_{k}}^{Bh}\rho \sigma
_{e_{k}g_{j}}^{Bh}+\rho \sigma _{e_{k}e_{k}}^{Bh}\right), \  \label{11}
\end{eqnarray}%
where $\gamma _{0j}^{A}$ and $\gamma _{kj}^{Bh}$ denote the spontaneous
radiation rates from state $\left\vert e_{0}\right\rangle _{A}$ to $\left\vert
g_{j}\right\rangle _{A}$ and $\left\vert e_{k}\right\rangle _{B}$ to $%
\left\vert g_{j}\right\rangle _{B}$ of the $h$th atom in the BEC, respectively; $\kappa _{ik}$ and $%
\kappa _{fk}$ denote the photon leakage rates from the cavity fields and fiber modes,
respectively; $\sigma _{mn}^{i}=\left\vert m\right\rangle _{i}\left\langle
n\right\vert $ ($m$, $n=e_{0}$, $e_{k}$, $g_{j}$) are the usual Pauli
matrices. Starting with the initial density matrix $\left\vert \phi
_{1}\right\rangle \left\langle \phi _{1}\right\vert $, by solving
numerically Eq. (11) in the subspace spanned by the vectors (5) and $%
\left\vert \phi _{13}\right\rangle =\left\vert g_{L}\right\rangle
_{A}\left\vert G_{0}\right\rangle _{B}\left\vert 0000\right\rangle
_{c}\left\vert 00\right\rangle _{f}$, $\left\vert \phi _{14}\right\rangle
=\left\vert g_{R}\right\rangle _{A}\left\vert G_{0}\right\rangle
_{B}\left\vert 0000\right\rangle _{c}\left\vert 00\right\rangle _{f}$. Fig. 3 shows the fidelity of the entanglement state as a function of the photon leakage rate $\kappa$ $\left( \kappa=\kappa
_{Ak}=\kappa _{Bk}=\kappa _{fk}\right)$
and for the atom spontaneous radiation
rate $\gamma $ $\left( \gamma =\sum_{j=a,L,R}\gamma
_{0j}^{A}=\sum_{j=L,0}\gamma
_{kj}^{Bh}=\sum_{j=R,0}\gamma _{kj}^{Bh}\right) $ = $0$, $0.2g$, $0.4g$, $0.6g$, $0.8g$, $1.0g$ (from the
top to the bottom).
\ In the calculation, for simplicity we choose $\gamma _{0a}^{A}=\gamma
_{0k}^{A}=\gamma /3$, $\gamma
_{k0}^{Bh}=\gamma _{kk}^{Bh}=\gamma /2$ $\left( k=l,r\right) $, the other parameters same as in
Fig. 2.  From the Fig. 3 we can see that the entanglement state
can be generated with highly fidelity even in the range of $\gamma $, $%
\kappa \sim g$.
\begin{figure}[htbp]
\centering\includegraphics[width=7cm]{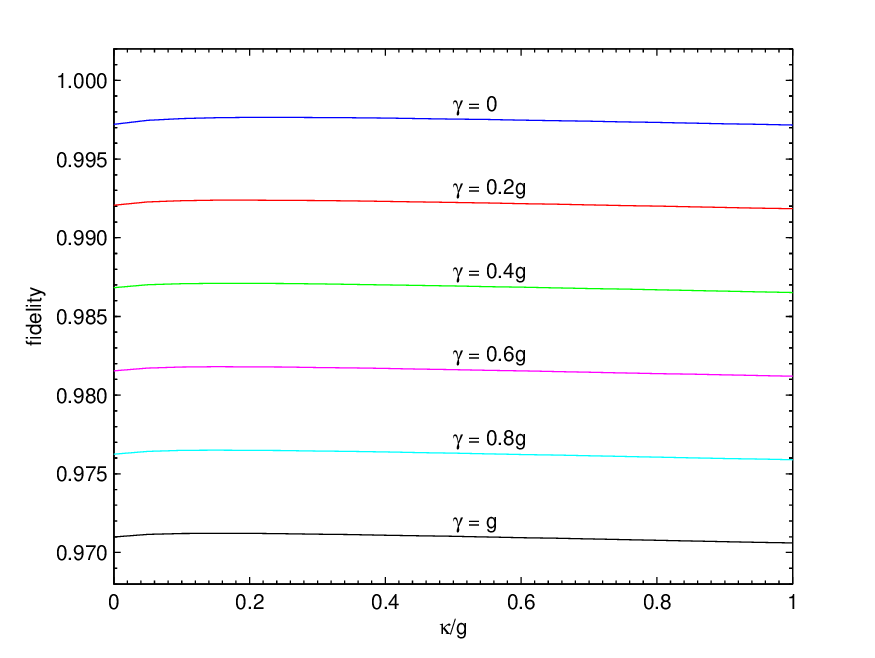}
\caption{Fidelity of the entanglement state (obtained by numerical simulation of master equation (8)) as a function of the photon leakage rate $\kappa$ and for the atom spontaneous radiation rate $\gamma=0$, $0.2g$, $0.4g$, $0.6g$, $0.8g$, $1.0g$ (from
the top to the bottom).}
\end{figure}
\section{ Discussion and conclusion}

It is necessary to briefly discuss the experimental feasibility of our
scheme. Firstly, trapping $^{87}Rb$\ BEC in cavity QED has also been realized in recently
experiment \cite{Brennecke}. In this experiment, the atom number can be
selected between $2,500$ and $200,000$ and the relevant cavity QED parameter
$(g,\kappa ,\gamma)=2\pi\times(10.6$, $1.3$, $3.0)$ MHz is realizable. So the
condition $\gamma$, $\kappa<0.4g$\ can be satisfied with these system
parameters for entangling the BEC and atom with fidelity larger than $98\%$. Secondly, the classical fields Rabi frequency can be selected by changing the laser density in principle. The strong coupling between two cavities by a waveguide has been experimental realized \cite{Sato}. The coupling strength can be reached as high as 25 GHz, which is much larger than atom-cavity coupling strength and the strength of the classical fields.
Finally, atoms in BEC do not fulfill the requirement of identical coupling, but it shows a similar energy spectrum, which can be modeled by the Tavis-Cummings Hamiltonian with an effective collective coupling $g_B^{eff} = g \mu \left( N \right)$, here $\mu\left(N\right) =\sqrt{0.5}\left( 1-0.0017N^{0.34}\right) $ is the overlap
between BEC spatial atomic mode and cavity mode \cite{Brennecke,Leslie}. So the coupling strength
$g_{B}(t)$ will decrease with increasing atom number $N$. We can increase the
$\Omega_{B}(t)$ accordingly to compensate this. One challenge here is photoassociation driven by the classical laser because it gradually reduces the BEC atom number N \cite{Lettner}. The fidelity
as a function of the atom number $N$ of the BEC is plotted in Fig. 4 with the
parameters $\gamma =\kappa=0.4g$, and the other parameters same as in Fig. 2. From the Fig. 2, we can see that our scheme is  robust to the variation of atom number in the BEC. Of course if the lost atoms carry away the single excitation, the scheme will be fail.
\begin{figure}[htbp]
\centering\includegraphics[width=7cm]{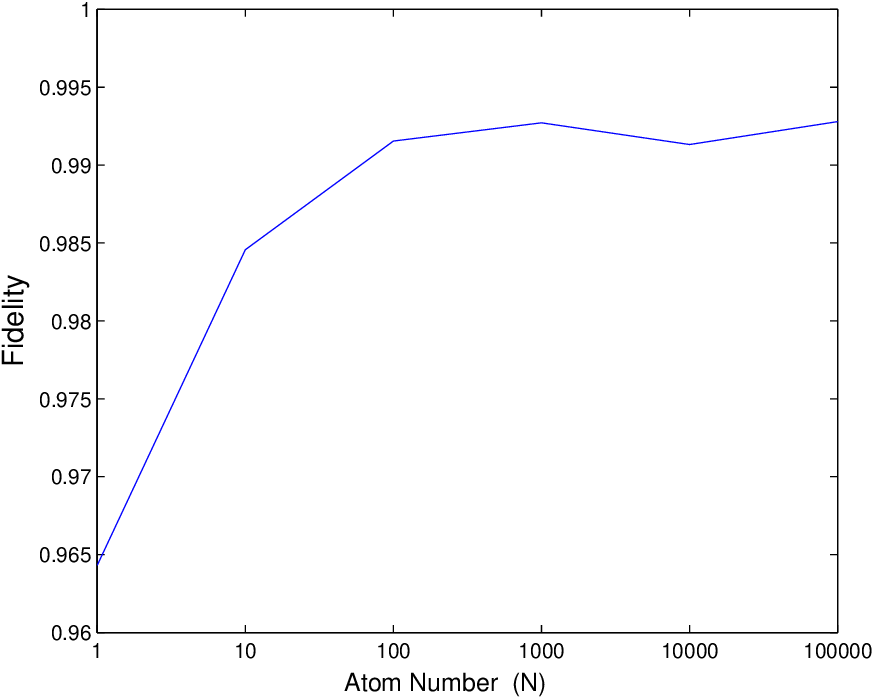}
\caption{Fidelity vs the atom number $N$ of the BEC with the parameters $%
\gamma=\kappa=0.4g$, and the other parameters same as in Fig. 2.}
\end{figure}

In summary, based on the STIRAP technique, we propose a scheme to prepare
three-dimensional entanglement state between a BEC and a atom. In this
scheme, the atomic spontaneous radiation and photon leakage can be efficiently
suppressed, since the populations of the excited states of atoms and cavity
(fiber) modes are almost zero in the whole process. We also show that this
scheme is highly stable to the variation of atom number in the BEC.
Recently, strong atom--field coupling for Bose--Einstein condensates in an
optical cavity on a chip \cite{Colombe} and strong coupling between distant
photonic nanocavities \cite{Sato} have been experimentally realized. So our
scheme is considered as a promising scheme for realizing entanglement
between BEC and atom on a photonic chip.
\section*{Acknowledgments}

This work was supported by the National Natural Science Foundation of China
(Grant No. 60677044, 11005099), the Fundamental Research Funds for the
central universities (Grant No. 201013037). L. Chen was also supported in part by the Government of China through CSC
(Grant No. 2009633075).

\end{document}